\documentclass[preprint,5p,sort&compress]{myelsarticle}
\graphicspath{{./figures/}{./build/figures/}{./code/out/figures/}}

\usepackage{pgfplots}
\pgfplotsset{compat=1.12}
\usepgfplotslibrary{groupplots}
\pgfplotsset{every axis/.append style={thick}}

\usepackage{makecell}

\usepackage{multirow}

\usepackage[labelfont=bf]{caption}
\usepackage{amsmath}
\numberwithin{equation}{section}
\usepackage{amssymb}
\usepackage[inline]{enumitem}
\usepackage[version=4]{mhchem}
\usepackage{siunitx}
\DeclareSIUnit\Molar{\textsc{m}}
\usepackage{url}
\newsavebox{\battery}

\newcommand{\pd}[2]{\frac{\partial #1}{\partial #2}}
\newcommand{\od}[2]{\frac{\mathrm{d}#1}{\mathrm{d}#2}}
\newcommand{\pds}[2]{\frac{\partial^2#1}{\partial #2^2}}

\newcommand{\bse}{\begin{subequations}}
\newcommand{\ese}{\end{subequations}}
\newcommand{\pow}[1]{^{(#1)}}
\newcommand{\myint}[4]{\int_{#1}^{#2} \! #3 \, \mathrm{d}#4}
\newcommand{\bm}[1]{\mathbf{#1}}
\newcommand{\di}[1]{\hat{#1}}
\newcommand{\Deff}{D^\text{eff}}
\newcommand{\Deffz}{D^{\text{eff},(0)}}
\newcommand{\keff}{\kappa^\text{eff}}
\newcommand{\keffz}{\kappa^{\text{eff},(0)}}

\newcommand{\Cd}{\mathcal{C}_\text{d}}
\newcommand{\jecd}{{j}_0}
\newcommand{\jecdn}{{j}_{0,\text{n}}}
\newcommand{\jecdp}{{j}_{0,\text{p}}}

\newcommand{\jecdnz}{{j}_{0,\text{n}}\pow{0}}
\newcommand{\jecdpz}{{j}_{0,\text{p}}\pow{0}}
\newcommand{\jecdno}{{j}_{0,\text{n}}\pow{1}}
\newcommand{\jecdpo}{{j}_{0,\text{p}}\pow{1}}

\newcommand{\Jecdn}{{J}_{0,\text{n}}}
\newcommand{\Jecdp}{{J}_{0,\text{p}}}
\newcommand{\Jecdnz}{{J}_{0,\text{n}}\pow{0}}
\newcommand{\Jecdpz}{{J}_{0,\text{p}}\pow{0}}
\newcommand{\Jecdno}{{J}_{0,\text{n}}\pow{1}}
\newcommand{\Jecdpo}{{J}_{0,\text{p}}\pow{1}}
\newcommand{\ocpn}{_{\ce{Pb}}}
\newcommand{\ocpp}{_{\ce{PbO_2}}}

\newcommand{\n}{_\text{n}}
\newcommand{\s}{_\text{sep}}
\newcommand{\p}{_\text{p}}
\newcommand{\elln}{\ell_\text{n}}

\newcommand{\ellp}{\ell_\text{p}}

\begin{document}

\title{Faster Lead-Acid Battery Simulations from Porous-Electrode Theory: \\
II. Asymptotic Analysis}

\author[mi]{Valentin~Sulzer\corref{cor1}}
\ead{sulzer@maths.ox.ac.uk}
\author[mi,far]{S.~Jon~Chapman}
\ead{chapman@maths.ox.ac.uk}
\author[mi,far]{Colin~P.~Please}
\ead{please@maths.ox.ac.uk}
\author[eng,far]{David~A.~Howey}
\ead{david.howey@eng.ox.ac.uk}
\author[eng,far]{Charles~W.~Monroe}
\ead{charles.monroe@eng.ox.ac.uk}

\cortext[cor1]{Corresponding author}
\address[mi]{Mathematical Institute, University of Oxford, OX2 6GG, United Kingdom}
\address[eng]{Department of Engineering Science, University of Oxford, OX1 3PJ, United Kingdom}
\address[far]{The Faraday Institution}

\begin{abstract}
Electrochemical and equivalent-circuit modelling are the two most popular approaches to battery simulation, but the former is computationally expensive and the latter provides limited physical insight. A theoretical middle ground would be useful to support battery management, on-line diagnostics, and cell design. We analyse a thermodynamically consistent, isothermal porous-electrode model of a discharging lead-acid battery. Asymptotic analysis of this full model produces three reduced-order models, which relate the electrical behaviour to microscopic material properties, but simulate discharge at speeds approaching an equivalent circuit. A lumped-parameter model, which neglects spatial property variations, proves accurate for C-rates below $0.1C$, while a spatially resolved higher-order solution retains accuracy up to $5C$. The problem of parameter estimation is addressed by fitting experimental data with the reduced-order models.
\end{abstract}
\maketitle

\section{Introduction}

The popular equivalent-circuit approach to battery modelling \cite{salameh1992mathematical} is efficient, but has limited physical detail and extrapolates poorly. Electrochemical models \cite{gu1987mathematical,alavyoon1991theoretical, bernardi1993two,
gu1997numerical, bernardi1995mathematical, newman1997simulation, gu2002modeling,
cugnet2011effect,boovaragavan2009mathematical}
require far more computational power, but include detailed descriptions of physical mechanisms, which presumably enhances predictive capability. Battery management could be improved if there existed easily-solved models with greater mechanistic detail. To that end, this paper puts forward several reduced-order models of lead-acid battery discharge, each derived from a mechanistic description based on an extension of Newman's porous-electrode theory \cite{newman2012electrochemical}, which we developed in part I.

Several authors have simplified mechanistic lead-acid-battery models to improve their computational efficiency. Newman and Tiedemann~\cite{newman1997simulation} recognise that  spatial gradients can be ignored at low current; they state a `lumped parameter model' (LPM) that depends only on time, but do not show how it derives from a porous-electrode model.
Gandhi \textit{et al.}~\cite{gandhi2009simplified} propose a LPM to underpin an analytical current/voltage relation.
Knauff~\cite{knauff2013kalman} simplifies a porous-electrode model by assuming, without justification, that current is linear in space, and acid molarity, quadratic.

We deploy perturbation methods~\cite{hinch1991perturbation} to produce a hierarchy of increasingly complex models. After nondimensionalization, a \emph{diffusional C-rate}, $\Cd$--the C-rate scaled with the diffusion time-scale---is found to control how simply the full model can be approximated. Three reduced-order models are derived, validated against the full model, and applied to experiments for parameter estimation.

A leading-order expansion in the diffusional C-rate produces a LPM of the Newman--Tiedemann type, found to be accurate for C-rates below $0.1C$.
The first-order expansion accounts for quasi-static spatial heterogeneity within the electrode sandwich. As well as improving the fit of the full model, this correction has a computationally efficient closed-form expression.
Finally, the first-order solution is improved by accounting for diffusion transients. This composite model includes just one linear partial differential equation, but matches the full model well up to $5C$.

\section{Dimensionless model}
\label{sec:nondim}

In part I, we proposed a general three-dimensional, thermodynamically consistent, isothermal porous-electrode model of a discharging lead-acid battery.
The detailed model was simplified slightly on the basis of dimensional analysis to allow solution in a one-dimensional setting.

After nondimensionalization, we obtained the following dimensionless system governing the electrolyte concentration $c$, porosity $\varepsilon$, current density ${i}$ and potential $\Phi$, electrode current density ${i}_\text{s}$ and potential $\Phi_\text{s}$, and interfacial current density $j$:
\bse\label{eq:summary}
\begin{align}
\pd{}{t}(\varepsilon c) &= \frac{1}{\Cd}\pd{}{x}\left(\Deff\pd{c}{x}\right) + sj, \label{eq:dcdt}\\
\pd{\varepsilon}{t} &= -\beta^\text{surf}j, \label{eq:depsdt}\\
\pd{i}{x} &= j, \label{eq:didx}\\
\Cd\,{i} &= \keff\left(\chi\pd{\ln(c)}{x} - \pd{\Phi}{x}\right), \label{eq:i}\\
\pd{i_{\text{s}}}{x} &= -j, \label{eq:disdx}\\
{i}_{\text{s}} &= -\iota_{\text{s}}\pd{\Phi_{\text{s}}}{x}, \label{eq:is}\\
j &= 2\jecd\sinh\left(\Phi_\text{s}-\Phi-U(c)\right) + \gamma_{\text{dl}}\pd{}{t}\left(\Phi_\text{s}-\Phi\right), \label{eq:j}
\end{align}
with boundary conditions
\begin{align}
\Phi_{\text{s}} = \pd{c}{x} = i = 0, \quad i_\text{s} = \mathrm{i}_\text{cell} \quad &\text{ at } x = 0, 1, \label{eq:BCs_collectors}\\
i_\text{s} = 0 \quad &\text{ at } x = \ell_\text{n}, 1-\ell_\text{p},\label{eq:BCs_separator}
\end{align}
and initial conditions
\begin{align}\label{eq:ICs}
c &= q^0, \\
{\varepsilon} &= \varepsilon^\text{max}-\varepsilon^\Delta(1-q^0), \\
{\Phi} &= - {U}\ocpn(q^0), \\
{\Phi}_{\text{s}} &= \begin{cases}
0, \quad &0<{x}<\ell_\text{n}\\
{U}\ocpp\left({c}^0\right) - {U}\ocpn\left({c}^0\right), \quad &1-\ell_\text{p}<{x}<1.
\end{cases}
\end{align}
Equation \eqref{eq:disdx} with the boundary conditions \eqref{eq:BCs_collectors} and \eqref{eq:BCs_separator} also implies the integral condition
\begin{equation}\label{eq:j_BC}
	\myint{0}{\elln}{{j}_\text{n}}{\di{x}}
	= -\myint{1-\ellp}{1}{{j}_\text{p}}{\di{x}}
	= i_\text{cell},
\end{equation}
\ese
where property values in the negative and positive electrode are designated with subscripts n and p, respectively. Typical values of the dimensionless parameters $\Cd$, $\iota_\text{s}$, $\beta^\text{surf}$, $\gamma_\text{dl}$, $\ell$, $s$, $q^0$, $\varepsilon^\text{max}$ and $\varepsilon^\Delta$ are given in Table~\ref{tab:dimless_params},
while concentration-dependent functions $D$, $\kappa$, $\chi$, $j_0$ and $U$ are given in Table~\ref{tab:functions}.
The dimensionless applied current is $i_\text{cell}(t) = I_\text{circuit}(t)/8A_\text{cs}$, where $I_\text{circuit}(t)$ is the applied current in the external circuit and $A_\text{cs}$ is the electrode cross-sectional area. We define $\bar{i}$ to be the maximum value of $i_\text{cell}(t)$ with respect to time.

The key parameter is the diffusional C-rate, $\Cd$, which is the C-rate as measured on the diffusion time-scale (or alternatively, the ratio of the applied current scale to the scale of the limiting current).

In the Results section, we will take $q^0$ to be unity (the battery starts from a fully charged state) unless explicitly stated.

\begin{table}[t]
\centering
\begin{tabular}{|c|c c c|}
\hline
\multirow{2}{*}{Parameter} & \multicolumn{3}{c|}{Value} \\
\cline{2-4}
& n & sep & p \\
\hline
$\Cd$ & \multicolumn{3}{c|}{$0.60\mathcal{C}$} \\
$\iota_\text{s}$ &  $3.8\times10^4/\mathcal{C}$ & - & $55/\mathcal{C}$ \\
$\beta^\text{surf}$ & $0.084$ & - & $-0.064$ \\
$\gamma_\text{dl}$ & $2.1\times10^{-5}$ & - & $1.7\times10^{-4}$ \\
$\ell$ & $0.25$ & $0.41$ & $0.34$ \\
$s$ & $-0.2$ & - & $0.8$ \\
$q^0$ & \multicolumn{3}{c|}{$1$} \\
${\varepsilon}^\text{max}$ & $0.53$ & $0.92$ & $0.57$ \\
${\varepsilon}^\Delta$ & $0.24$ & - & $-0.13$ \\
\hline
\end{tabular}
\caption{Dimensionless parameters, relative to the C-rate, ${\mathcal{C} = I_\text{circuit}/Q}$. Further details and interpretations can be found in part I.}
\label{tab:dimless_params}
\end{table}

\section{Solutions}
\label{sec:solutions}

We now derive three analytical, approximate solutions to the model system \eqref{eq:summary}, and compare these to the numerical solution of the full model computed in part I, which we treat as `ground truth'. To do this, we note that the diffusional C-rate, $\Cd$, is small for most practical (low C-rate) applications, and perform an asymptotic analysis near the limit of small $\Cd$.

\subsection{Leading-order quasi-static solution}
\label{sec:sol_O1}

In this section, we will derive the quasi-static solution in the limit of small $\Cd$, $\gamma_\text{dl}$ and $1/\iota_\text{s}$. Since $\gamma_{\text{dl}}$ and $1/\iota_\text{s}$ are much smaller than one (Table~\ref{tab:dimless_params}), we only take the leading-order terms in their expansions. In contrast, $\Cd$ can sometimes be close to one, so we will consider both the leading order and first order in $\Cd$.

To leading order in $1/\iota_\text{s}$, \eqref{eq:is} becomes
\begin{equation}\label{eq:Phis_uniform}
\pd{\Phi_{\text{s}}}{x} = 0
\end{equation}
in each electrode, and so $\Phi_\text{s}$ can be approximated as a function of time only. Applying the boundary condition \eqref{eq:BCs_collectors} and defining $V(t) = \left.\Phi_\text{s}\right\rvert_{x=1}$, we can now replace \eqref{eq:is} with
\begin{equation}\label{eq:Phis_largeiotas}
\Phi_{\text{s},\text{n}} = 0, \qquad \Phi_{\text{s},\text{p}} = V(t).
\end{equation}
We use the integral condition \eqref{eq:j_BC} so that we do not need to solve for $i_\text{s}$ to find the voltage, $V(t)$. Hence \eqref{eq:is} is only necessary if we want to find $i_\text{s}$ having found $j$.
We also take the leading order in $\gamma_\text{dl}$, so that the time derivatives in \eqref{eq:j} disappear.

In summary, we simplify the system \eqref{eq:summary} to the following equations for $c(x,t)$, $\varepsilon(x,t)$, $j(x,t)$, $\Phi(x,t)$ and $V(t)$:
\bse\label{eq:summary_simplified}
\begin{align}
\pd{}{t}(\varepsilon c) &= \frac{1}{\Cd}\pd{}{x}\left(\Deff\pd{c}{x}\right) + sj, \label{eq:dcdt_simplified}\\
\pd{\varepsilon}{t} &= -\beta^\text{surf}j, \label{eq:depsdt_simplified}\\
\Cd\,{j} &= \pd{}{x}\left[\keff\left(\chi\pd{\ln(c)}{x} - \pd{\Phi}{x}\right)\right], \label{eq:i_simplified}\\
j_\text{n} &= 2\jecdn\sinh\left(-\Phi-U\ocpn(c)\right), \label{eq:jn_simplified} \\
j_\text{p} &= 2\jecdp\sinh\left(V-\Phi-U\ocpp(c)\right),\label{eq:jp_simplified}
\end{align}
with boundary conditions
\begin{align}
&\pd{c}{x} = \pd{\Phi}{x} = 0 \quad \text{ at } x = 0, 1,\\
&\myint{0}{\ell_\text{n}}{j_\text{n}}{x} = -\myint{1-\ell_\text{p}}{1}{j_\text{p}}{x} = \mathrm{i}_\text{cell},
\end{align}
and initial conditions \eqref{eq:ICs}.
\ese

As shown in Table~\ref{tab:dimless_params}, the diffusional C-rate, $\Cd$, is equal to $0.6\mathcal{C}$, where $\mathcal{C}$ is the C-rate. Most practical applications have a C-rate below 0.25C, so the diffusional C-rate is usually small. Hence we perform an asymptotic expansion in the limit $\Cd\to0$ and assume that we can expand all variables in \eqref{eq:summary_simplified} in powers of $\Cd$:
\begin{equation}\label{eq:Da_expansion}
f(x,t) = f\pow{0}(x,t) + \Cd f\pow{1}(x,t)+ \Cd^2 f\pow{2}(x,t) + \mathcal{O}(\Cd^3),
\end{equation}
where $f = c, \varepsilon, \Phi, j$ and $V$. Hence \eqref{eq:summary_simplified} becomes to leading order
\bse\label{eq:summary_O1}
\begin{align}
0 &= \pd{}{x}\left(D^{\text{eff},(0)}\pd{c\pow{0}}{x}\right), \label{eq:dcdt_O1}\\
\pd{\varepsilon\pow{0}}{t} &= -\beta^\text{surf}j\pow{0}, \label{eq:depsdt_O1}\\
0 &= \pd{}{x}\left[\keffz\left(\chi\pow{0}\pd{\ln\left(c\pow{0}\right)}{x} - \pd{\Phi\pow{0}}{x}\right)\right], \label{eq:i_O1}\\
j\pow{0}_\text{n} &= 2\jecdnz\sinh\left(-\Phi\pow{0}-U\ocpn\left(c\pow{0}\right)\right)\label{eq:jn_O1}\\
j\pow{0}_\text{p} &= 2\jecdpz\sinh\left(V\pow{0}-\Phi\pow{0}-U\ocpp\left(c\pow{0}\right)\right).\label{eq:jp_O1}
\end{align}
The leading-order diffusivity is
${\Deffz = D\left(c\pow{0}\right)\left(\varepsilon\pow{0}\right)^{3/2}}$,
and similarly for $\keffz$, $\chi\pow{0}$ and $c_\text{w}\pow{0}$,
The boundary conditions are
\begin{align}
&\pd{c\pow{0}}{x} = \pd{\Phi\pow{0}}{x} = 0 \quad \text{ at } x = 0, 1, \label{eq:BCs_collectors_O1}\\
&\myint{0}{\ell_\text{n}}{j\pow{0}_\text{n}}{x} = -\myint{1-\ell_\text{p}}{1}{j\pow{0}_\text{p}}{x} = \mathrm{i}_\text{cell},\label{eq:j_BC_O1}
\end{align}
and the initial conditions are
\begin{align}\label{eq:ICs_O1}
&c\pow{0} = c^0, \qquad {\varepsilon}\pow{0} = \varepsilon^0.
\end{align}
\ese
At first order, equating coefficients of $\Cd$ in \eqref{eq:summary_simplified} gives
\bse\label{eq:summary_ODa}
\begin{align}
\pd{}{t}&\left(\varepsilon\pow{0}c\pow{0}\right) = \pd{}{x}\left(D^{\text{eff},(0)}\pd{c\pow{1}}{x}\right) + sj\pow{0}, \label{eq:dcdt_ODa}\\
\pd{\varepsilon\pow{1}}{t} &= -\beta^\text{surf}j\pow{1}, \label{eq:depsdt_ODa}\\
j\pow{0} &= \pd{}{x}\left[\keffz\left(\frac{\chi\pow{0}}{c\pow{0}}\pd{c\pow{1}}{x} - \pd{\Phi\pow{1}}{x}\right)\right], \label{eq:i_ODa}\\
j\pow{1}_\text{n} &= 2\left(\jecdno\sinh\left[{\eta}\pow{0}_\text{n}\right]+ \jecdnz\eta\pow{1}_\text{n}\cosh\left[{\eta}\pow{0}_\text{n}\right]\right), \label{eq:jn_ODa}\\
j\pow{1}_\text{p} &= 2\left(\jecdpo\sinh\left[{\eta}\pow{0}_\text{p}\right] + \jecdpz\eta\pow{1}_\text{p}\cosh\left[{\eta}\pow{0}_\text{p}\right]\right),\label{eq:jp_ODa}
\end{align}
where
\begin{align}
&\eta\pow{0}_\text{n} = -\left({\Phi}\pow{0} + {U}\ocpn(c\pow{0})\right), \\
&\eta\pow{0}_\text{p} = V\pow{0} - {\Phi}\pow{0} - {U}\ocpp\left(c\pow{0}\right), \\
&\eta\pow{1}_\text{n} = -\left({\Phi}\pow{1} + c\pow{1}{U}'\ocpn\left(c\pow{0}\right)\right), \\
&\eta\pow{1}_\text{p} = V\pow{1} - {\Phi}\pow{1} - c\pow{1}{U}'\ocpp\left(c\pow{0}\right),
\end{align}
with boundary conditions
\begin{align}
&\pd{c\pow{1}}{x} = \pd{\Phi\pow{1}}{x} = 0 \quad \text{ at } x = 0, 1, \label{eq:BCs_collectors_ODa}\\
&\myint{0}{\ell_\text{n}}{j\pow{1}}{x} = -\myint{1-\ell_\text{p}}{1}{j\pow{1}}{x} = 0, \label{eq:j_BC_ODa}
\end{align}
and initial conditions
\begin{align}\label{eq:ICs_ODa}
c\pow{1} = {\varepsilon}\pow{1} = 0&\quad \text{ at $t=0$}.
\end{align}
\ese

\paragraph{Leading-order quasi-static solution}
We now seek the solution to the lowest order problem.
Integrating \eqref{eq:dcdt_O1} with boundary conditions \eqref{eq:BCs_collectors_O1}, then integrating again, gives $c\pow{0} = c\pow{0}(t)$. We then integrate \eqref{eq:i_O1}, use boundary conditions \eqref{eq:BCs_collectors_O1}, and integrate again, to find that $\Phi\pow{0} = \Phi\pow{0}(t)$. Hence $j\pow{0}_\text{n}$ and $j\pow{0}_\text{p}$ as defined by \eqref{eq:jn_O1} and \eqref{eq:jp_O1} are functions of time only; the boundary conditions \eqref{eq:j_BC_O1} give
\begin{align}\label{eq:j0}
j\pow{0}_\text{n} = \mathrm{i}_\text{cell}/\ell_\text{n}, \qquad j\pow{0}_\text{p} = -\mathrm{i}_\text{cell}/\ell_\text{p}.
\end{align}
Finally, by \eqref{eq:depsdt_O1}, $\varepsilon_\text{n}\pow{0}$, $\varepsilon_\text{sep}\pow{0}$ and $\varepsilon_\text{p}\pow{0}$ are functions of time only (in particular, $\varepsilon_\text{sep}\pow{0} \equiv \varepsilon_\text{sep}^\text{max}$). Hence to leading order, the whole problem is quasi-static.
To determine $c\pow{0}$, we need to consider the first-order problem \eqref{eq:dcdt_ODa} for $c\pow{1}$.
Integrating \eqref{eq:dcdt_ODa} from $x=0$ to $x=1$ and using \eqref{eq:j0} and the boundary conditions \eqref{eq:BCs_collectors_ODa} gives a solvability condition that determines $c\pow{0}$. We can combine this with \eqref{eq:depsdt_O1}, \eqref{eq:jn_O1} and \eqref{eq:jp_O1} to obtain a nonlinear differential-algebraic equation system governing $c\pow{0}$, $\varepsilon_\text{n}\pow{0}$, $\varepsilon_\text{p}\pow{0}$, $\Phi\pow{0}$ and $V\pow{0}$,
\bse\label{eq:O1_ODEs}
\begin{align}
	\od{c\pow{0}}{t} &= \frac{1}{\ell_\text{n}\varepsilon_\text{n}\pow{0} + \ell_\text{sep}\varepsilon_\text{sep}^\text{max} + \ell_\text{p}\varepsilon_\text{p}\pow{0}}\nonumber
	\\
	&\times\left[(s_\text{n}-s_\text{p})\mathrm{i}_\text{cell}\phantom{\pd{}{}}\right.\nonumber
	\\
	& \left.- c\pow{0}\od{}{t}\left(\ell_\text{n}\varepsilon_\text{n}\pow{0} + \ell_\text{sep}\varepsilon_\text{sep}^\text{max} + \ell_\text{p}\varepsilon_\text{p}\pow{0}\right)\right],
	\\
	\od{\varepsilon_\text{n}\pow{0}}{t} &= -\frac{\beta^\text{surf}_\text{n}\mathrm{i}_\text{cell}}{\ell_\text{n}},
	\\
	\od{\varepsilon_\text{p}\pow{0}}{t} &= \frac{\beta^\text{surf}_\text{p}\mathrm{i}_\text{cell}}{\ell_\text{p}},
	\\
	\mathrm{i}_\text{cell}/\ell_\text{n} &= 2\jecdnz\sinh\left(-\Phi\pow{0}-U\ocpn\left(c\pow{0}\right)\right),
	\\
	-\mathrm{i}_\text{cell}/\ell_\text{p} &= 2\jecdpz\sinh\left(V\pow{0}-\Phi\pow{0}-U\ocpp\left(c\pow{0}\right)\right),
\end{align}
\ese
with initial conditions \eqref{eq:ICs_O1}. Integrate (\ref{eq:O1_ODEs}a-c) and rearrange (\ref{eq:O1_ODEs}d,e) to find the final leading-order solution,
\bse\label{eq:O1_algebraic}
\begin{align}
&c\pow{0} = \frac{\left(\ell_\text{n}\varepsilon_{\text{n}}^0 + \ell_\text{sep}\varepsilon_\text{sep}^\text{max} + \ell_\text{p}\varepsilon_{\text{p}}^0\right)q^0 + (s_\text{n}-s_\text{p})\myint{0}{t}{\mathrm{i}_\text{cell}}{s}}{\ell_\text{n}\varepsilon_\text{n}\pow{0} + \ell_\text{sep}\varepsilon_\text{sep}^\text{max} + \ell_\text{p}\varepsilon_\text{p}\pow{0}}, \label{eq:c0}\\
&\varepsilon_\text{n}\pow{0} = \varepsilon^0_\text{n} -\frac{\beta^\text{surf}_\text{n}}{\ell_\text{n}}\myint{0}{t}{\mathrm{i}_\text{cell}}{s}, \\
&\varepsilon_\text{p}\pow{0} = \varepsilon^0_\text{p} + \frac{\beta^\text{surf}_\text{p}}{\ell_\text{p}}\myint{0}{t}{\mathrm{i}_\text{cell}}{s}, \\
&\Phi\pow{0} = - U\ocpn\left(c\pow{0}\right) - \sinh^{-1}\left(\frac{\mathrm{i}_\text{cell}}{2\jecdnz\ell_\text{n}}\right), \\
&V\pow{0} = U\ocpp\left(c\pow{0}\right) - U\ocpn\left(c\pow{0}\right) \nonumber\\
&{\phantom{V\pow{0}=} } - \sinh^{-1}\left(\frac{\mathrm{i}_\text{cell}}{2\jecdnz\ell_\text{n}}\right) - \sinh^{-1}\left(\frac{\mathrm{i}_\text{cell}}{2\jecdpz\ell_\text{p}}\right).\label{eq:V0}
\end{align}
\ese

\subsection{First-order quasi-static solution}
\label{sec:sol_ODa}

We now solve the first-order system, \eqref{eq:summary_ODa}, to find the $\mathcal{O}(\Cd)$ correction to the voltage. We solve \eqref{eq:summary_ODa} as follows:
\begin{enumerate*}[label=(\roman*)]
\item find $c\pow{1}$ using \eqref{eq:dcdt_ODa}, up to an arbitrary constant, $k(t)$;
\item find $k$ using a solvability condition on $c\pow{2}$, the $\mathcal{O}(\Cd^2)$ correction to $c$;
\item find $\Phi\pow{1}$ using \eqref{eq:i_ODa} up to an arbitrary constant, $A_\text{n}(t)$;
\item find $A_\text{n}$ using \eqref{eq:jn_ODa};
\item find $V\pow{1}$ using \eqref{eq:jp_ODa}.
\end{enumerate*}
Firstly, with known $c\pow{0}$ and $\varepsilon\pow{0}$, we can integrate \eqref{eq:dcdt_ODa} with respect to $x$ twice and use \eqref{eq:BCs_collectors_ODa} to find an explicit equation for $c\pow{1}$ (given in \ref{app:c1}).

Having found $c\pow{1}$, we integrate \eqref{eq:i_ODa}, using \eqref{eq:j0} and continuity of $\Phi\pow{1}$, to find
\begin{equation}\label{eq:Phi1}
\begin{aligned}
\Phi\pow{1} = &\frac{\chi\pow{0}c\pow{1}}{c\pow{0}}+ A_\text{n} \\
&- \begin{cases}
\frac{\mathrm{i}_\text{cell}x^2}{2\ell_\text{n}\keffz_\text{n}} , \quad& 0<x<\ell_\text{n}, \\
{\mathrm{i}_\text{cell}}\left(\frac{\ell_\text{n}}{2\keffz_\text{n}} + \frac{x-\ell_\text{n}}{\keffz_\text{s}}\right), \quad& \ell_\text{n}<x<1 - \ell_\text{p}, \\
\mathrm{i}_\text{cell}\left(\frac{\ell_\text{n}}{2\keffz_\text{n}} + \frac{\ell_\text{sep}}{\keffz_\text{s}}\right.\\
\left. \hspace{1cm}+ \frac{\ell_\text{p}^2 - (1-x)^2}{2\ell_\text{p}\keffz_\text{p}}\right), \quad& 1 - \ell_\text{p}<x<1,
\end{cases}
\end{aligned}
\end{equation}
where $A_\text{n}$ is an arbitrary constant.

We can now integrate \eqref{eq:jn_ODa} from $x=0$ to $x=\ell_\text{n}$ and integrate \eqref{eq:jp_ODa} from $x=1-\ell_\text{p}$ to $x=1$, using \eqref{eq:j_BC_ODa} each time, to find the correction term $V\pow{1}$:
\bse\label{eq:ODa_algebraic}
\begin{multline}
A_\text{n} = \frac{\bar{j}_{0,\text{n}}\pow{1}\tanh\left({\eta}_\text{n}\pow{0}\right)}{\jecdnz} - \bar{c}_\text{n}\pow{1}{U}'\ocpn\left(c\pow{0}\right) \\
- \frac{\chi\pow{0}\bar{c}_\text{n}\pow{1}}{c\pow{0}} + \frac{\mathrm{i}_\text{cell}\ell_\text{n}}{6\keffz_\text{n}},
\end{multline}
\begin{equation}
V\pow{1} =  \bar{\Phi}\pow{1}_\text{p} + \bar{c}_\text{p}\pow{1}{U}'\ocpp\left(c\pow{0}\right) - \frac{\bar{j}_{0,p}\pow{1}\tanh\left({\eta}_\text{p}\pow{0}\right)}{\jecdpz},
\end{equation}
\ese
where we have introduced the averages
\begin{equation}
\bar{\cdot}_\text{n} = \frac{1}{\ell_\text{n}}\myint{0}{\ell_\text{n}}{\cdot}{x}, \qquad \bar{\cdot}_\text{p} = \frac{1}{\ell_\text{p}}\myint{1-\ell_\text{p}}{1}{\cdot}{x}.
\end{equation}

\subsection{Composite solution}
\label{sec:sol_comp}

The quasi-static solution developed in the `Leading-order quasi-static solution' and `First-order quasi-static solution' sections is valid when the current varies slowly, but fails to capture transient behaviour when the current changes more rapidly, such as a jump. To capture such transients, we could rescale time with $\tau = (t-t^*)/\Cd$, where $t^*$ is the time of the jump in the current, define $C(\tau) = c(t)$ (and likewise for other variables) and expand in powers of $\Cd$. We give the details of such an approach in \ref{app:transient}.

Such a transient solution is valid at short times after a jump time $t^*$, but breaks down at times long after the jump time.
To obtain a solution that is valid both at short times after a jump in current and at long times, without having to repeatedly `reset' the transient solution, we use a `composite' solution, which we now develop here.

We consider the lowest order and first order correction for the concentration by taking $\tilde{c} = c\pow{0} + \Cd\,c\pow{1}$. We then consider the PDE
\begin{equation}\label{eq:c_comp}
\varepsilon\pow{0}\pd{\tilde{c}}{t} = \frac{\Deffz}{\Cd}\pds{\tilde{c}}{x} + \left(s + \beta^\text{surf}c\pow{0}\right)j\pow{0},
\end{equation}
where $c\pow{0}$ and $\varepsilon\pow{0}$ are given by the quasi-static problem \eqref{eq:O1_algebraic} and $j\pow{0}$ is given by \eqref{eq:j0}. We note that
for long times, $\Cd\,\partial c\pow{1}/\partial t$ is a higher-order term and we retrieve the quasi-static problem \eqref{eq:dcdt_ODa}, while for short times, re-scaling $\tau = (t-t^*)/\Cd$, $c\pow{0}$ is constant and we have the transient problem \eqref{eq:transient_ODa_c} for $c\pow{1}$.
Hence \eqref{eq:c_comp} is valid uniformly at both short times and long times.

The composite solution then consists of solving \eqref{eq:c_comp} for $\tilde{c}$, then computing
\begin{equation}\label{eq:c1_comp}
c\pow{1} = \frac{\tilde{c}-c\pow{0}}{\Cd},
\end{equation}
and finally finding $V\pow{1}$ through \eqref{eq:Phi1} and \eqref{eq:ODa_algebraic} with $c\pow{1}$ given by \eqref{eq:c1_comp}.

\section{Results}
\label{sec:results}
\begin{figure*}[t]
	\centering
  \includegraphics{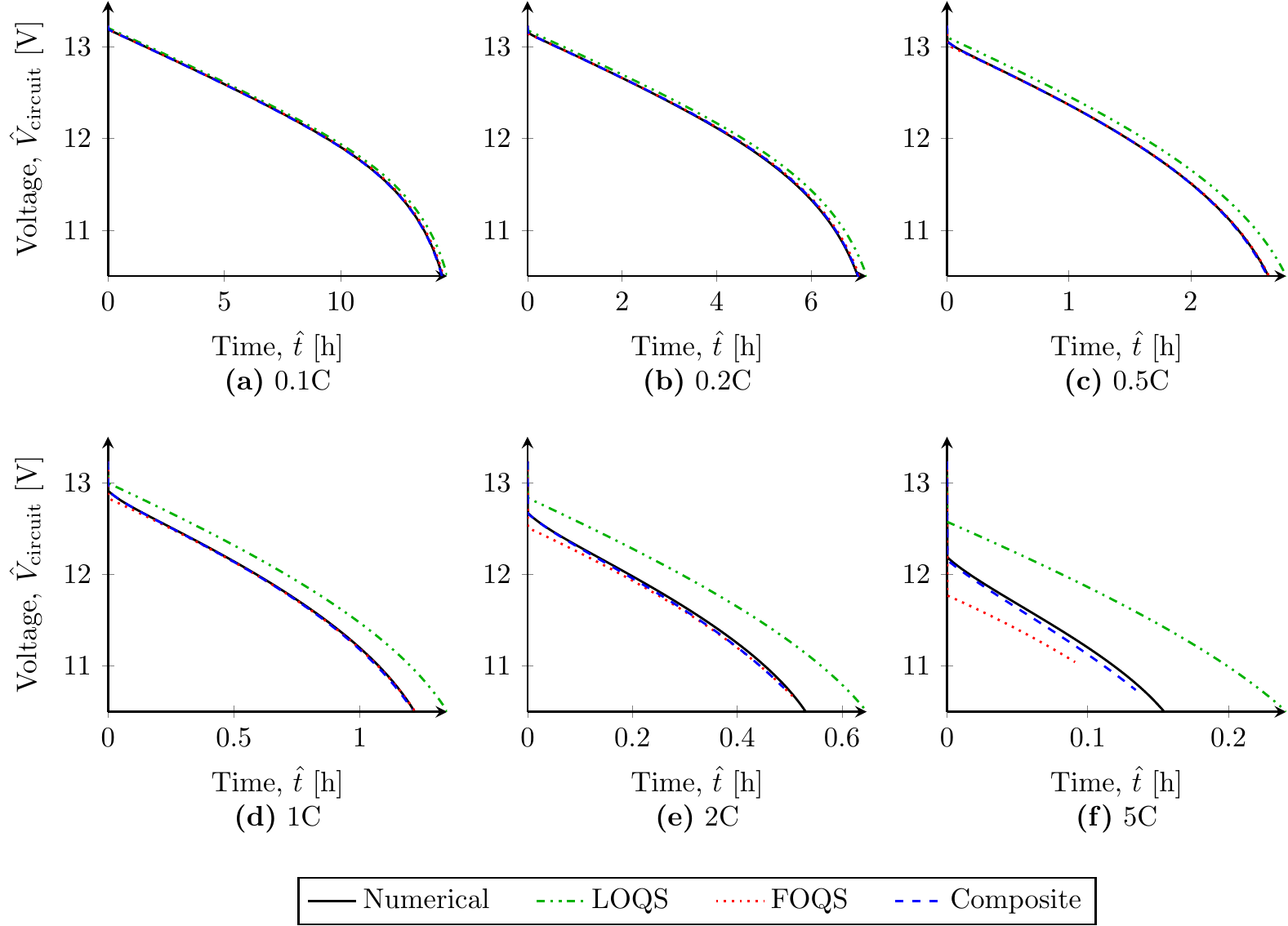}
	\caption{Comparing voltages for a constant-current discharge using the parameters from literature (Tables~\ref{tab:dimless_params} and \ref{tab:dim_params}), for a range of C-rates.}
	\label{fig:compare_voltages}
\end{figure*}

In the Solutions section,
we derived four systems that are approximately equivalent to the full dimensionless system \eqref{eq:summary} with varying degrees of accuracy:
\begin{enumerate}
  \item Numerical 
    -- part I
  \item Leading-order quasi-static (LOQS) 
    -- \eqref{eq:O1_algebraic}
  \item First-order quasi-static (FOQS) 
    -- \eqref{eq:c1}, \eqref{eq:Phi1} and \eqref{eq:ODa_algebraic}
  \item Composite 
    -- \eqref{eq:c_comp}, \eqref{eq:Phi1} and \eqref{eq:ODa_algebraic}
\end{enumerate}
The code used to solve the models and generate the results below is available publicly on GitHub \cite{valentin_sulzer_2019_2554000}.
Note that to obtain either the first-order quasi-static solution or the composite solution, we must first solve the leading-order quasi-static problem.

\subsection{Reduced-order solutions}

We now compare results from the four models. We treat the full numerical model  as `ground truth', and investigate the speed and accuracy of the three other models compared to the numerical model.

The most important output from the model is the voltage, since this is the variable that we can compare to experimental data (treating current as a known input). In Figure~\ref{fig:compare_voltages}, we compare the voltage during a complete constant-current discharge at a range of C-rates. The discharge is deemed to be finished either when the concentration reaches zero anywhere in the cell, or when the voltage reaches a cut-off voltage of $10.5$V.

We observe that all three reduced-order solutions agree well with the numerical solution at very low C-rates (Figure~\ref{fig:compare_voltages}a). As we increase the C-rate (Figures\ref{fig:compare_voltages}b-d), only the first-order solutions (FOQS and composite) agree with the numerical solution; further, a discrepancy appears between the FOQS solution and the numerical solution at early times. Finally, for very high C-rates (Figures~\ref{fig:compare_voltages}e-f) the composite solution still agrees very well with the numerical solution, but the FOQS solution does not, and terminates early, for reasons that we explain below.

\begin{figure*}[t]
	\centering
  \includegraphics{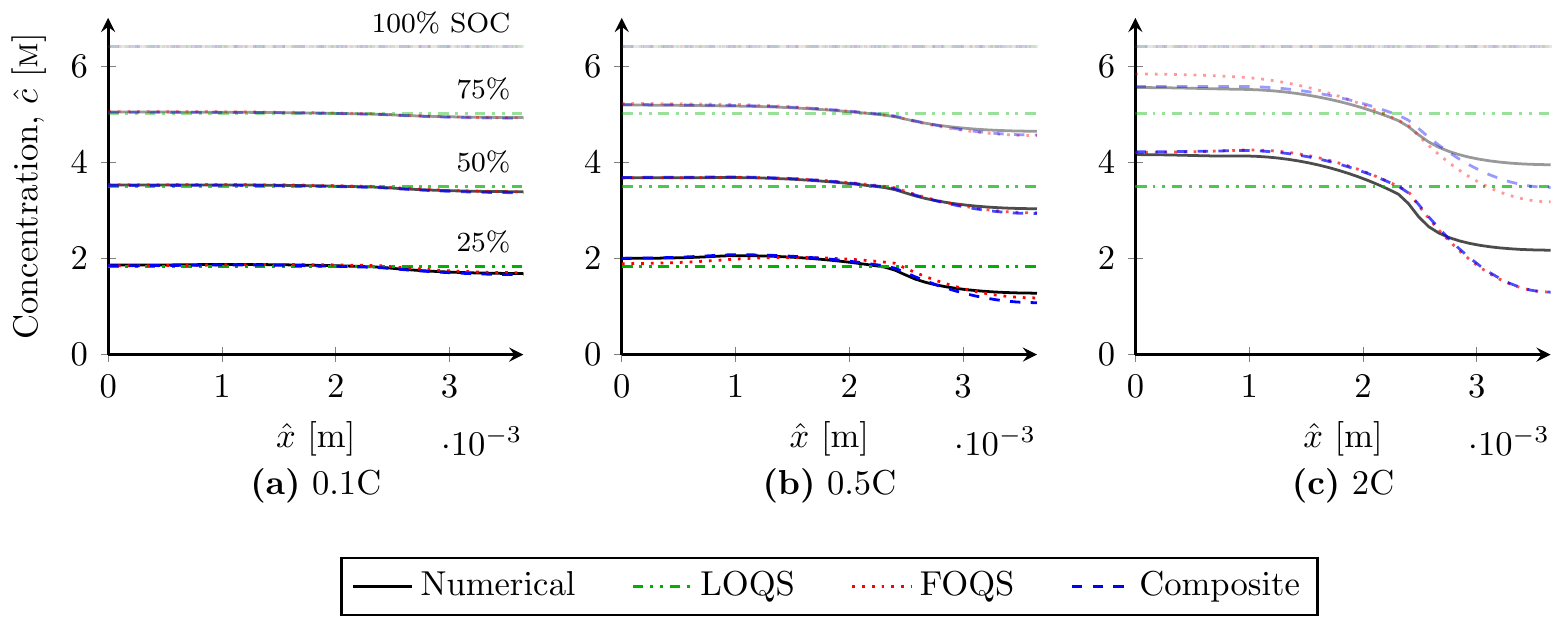}
	\caption{Comparing concentrations at various States of Charge (SOCs) for a constant-current discharge using the parameters from literature (Tables~\ref{tab:dimless_params} and \ref{tab:dim_params}), for a range of C-rates. Opacity increases with decreasing SOC. In (c), we only show the curves down to 50\% SOC, as the numerical, composite and LOQS solutions terminate before 25\% SOC.}
	\label{fig:compare_concentrations}
\end{figure*}

\def \Iinternala {0.1C}
\def \Iinternalb {0.5C}
\def \Iinternalc {2C}
To explain the behaviour observed in the voltages, we investigate internal variables, such as the concentration at various states of charge (Figure~\ref{fig:compare_concentrations}). At a very low C-rate of \Iinternala~(Figure~\ref{fig:compare_concentrations}a), the concentration remains almost uniform throughout the discharge; hence the LOQS solution, which does not take into account any spatial variations, provides a good fit to the numerical solution. At a higher C-rate of \Iinternalb~(Figure~\ref{fig:compare_concentrations}b), the concentration in the numerical solution is no longer spatially homogeneous; this is non-uniformity is captured well by the FOQS and composite solutions, but not by the LOQS solution. However, even with the FOQS and composite solutions, there is a discrepancy in the concentration profiles in the positive electrode (Figures~\ref{fig:compare_concentrations}b,c, right-hand side of the spatial domain). This is because the solutions from the asymptotic methods assume a uniform interfacial current density, but in the numerical solution the interfacial current density is non-uniform.

Finally, at high C-rates (\Iinternalc, Figure~\ref{fig:compare_concentrations}c), there is a diffusion transient at the start of the discharge; this is only captured by the composite solution, and not the FOQS solution. This initial diffusion transient also explains the discrepancy between the FOQS and numerical solutions at early times in Figure~\ref{fig:compare_voltages}d. In addition to this, we can now see that the FOQS solution terminates early in Figure~\ref{fig:compare_voltages}f because the concentration quickly reaches zero.

\begin{figure}
	\centering
  \includegraphics{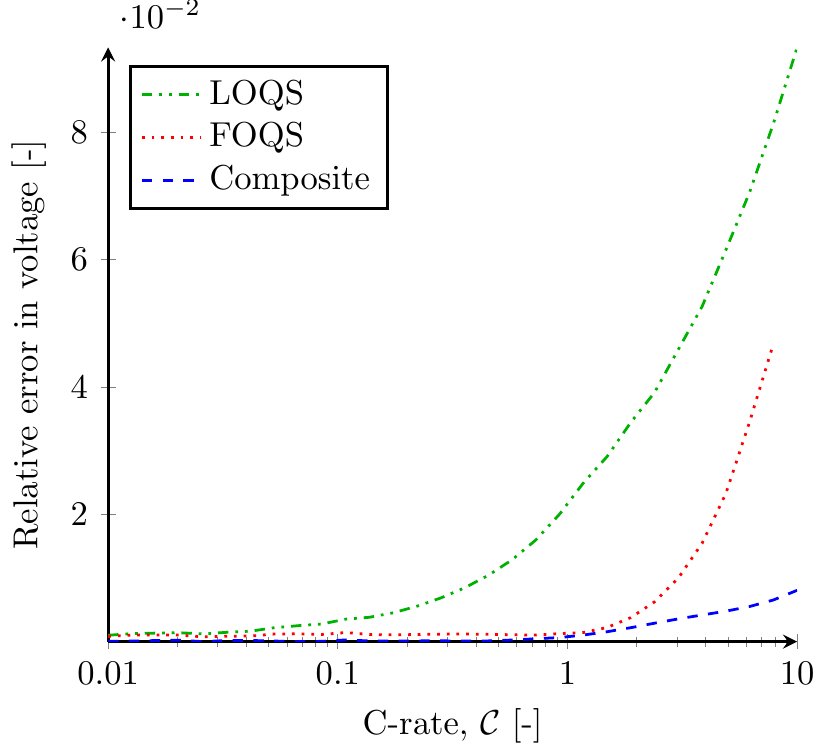}
	\caption{Relative error of the reduced-order models compared to the numerical model, for a constant-current discharge using the parameters from literature (Tables~\ref{tab:dimless_params} and \ref{tab:dim_params}), for a range of currents.}
	\label{fig:errors}
\end{figure}

In Figure~\ref{fig:errors}, we show the relative errors of the voltage obtained from reduced-order models compared to the voltage obtained from the numerical model.
Then, in Table~\ref{tab:speed}, we compare the time taken to solve the various models.
We see that the composite solution, leading-order quasi-static and first-order quasi-static solutions are roughly one, two and three orders of magnitude faster than the full numerical solution respectively. Coupled with the errors shown in Figure~\ref{fig:errors}, the speeds shown in Table~\ref{tab:speed} suggest that in order to solve the model accurately and as quickly as possible, we should use the LOQS model for C-rates below 0.1C, the FOQS model for C-rates of 0.1-1C, and the composite model for C-rates above 1C.

The time taken for the leading-order quasi-static model is independent of grid size, while the time taken for the other models scales linearly with grid size.
Note that we can expect to obtain a faster numerical solution by using a different spatial discretisation scheme than Finite Volumes, such as Chebyshev orthogonal collocation~\cite{bizeray2015lithium}, and the relative speed-up of the composite solution by using the same discretisation would be similar.
\renewcommand{\arraystretch}{1}
\begin{table*}[t]
\centering
\begin{tabular}{|c|cc|cc|cc|cc|}
	\hline
	& \multicolumn{2}{c|}{0.1C}& \multicolumn{2}{c|}{0.5C}& \multicolumn{2}{c|}{2C}& \multicolumn{2}{c|}{5C}\\
	\cline{2-9}
	Solution Method & Time & Speed-up & Time & Speed-up & Time & Speed-up & Time & Speed-up \\
	\hline
	Numerical & 0.643 & -  & 0.265 & -  & 1.3 & -  & 1.19 & - \\
	Composite & 0.050 & 13  & 0.048 & 5  & 0.051 & 26  & 0.292 & 4 \\
	FOQS & 0.005 & 126  & 0.005 & 55  & 0.005 & 261  & 0.005 & 239 \\
	LOQS & 0.002 & 407  & 0.001 & 183  & 0.001 & 887  & 0.001 & 805 \\
	\hline
\end{tabular}
\caption{Speed comparison for the four models for a constant-current discharge using the parameters from literature (Tables~\ref{tab:dimless_params} and \ref{tab:dim_params}). Time is CPU time in seconds, obtained on an AMD FX(tm)-4350 Quad-Core Processor, and averaged over 100 runs; speed-up is ratio of numerical time to own time.}
\label{tab:speed}
\end{table*}

\subsection{Voltage breakdown}

\def \Ibreakdowna {0.1C}
\def \Ibreakdownb {0.5C}
\def \Ibreakdownc {5C}
\begin{figure*}[t]
	\centering
  \includegraphics{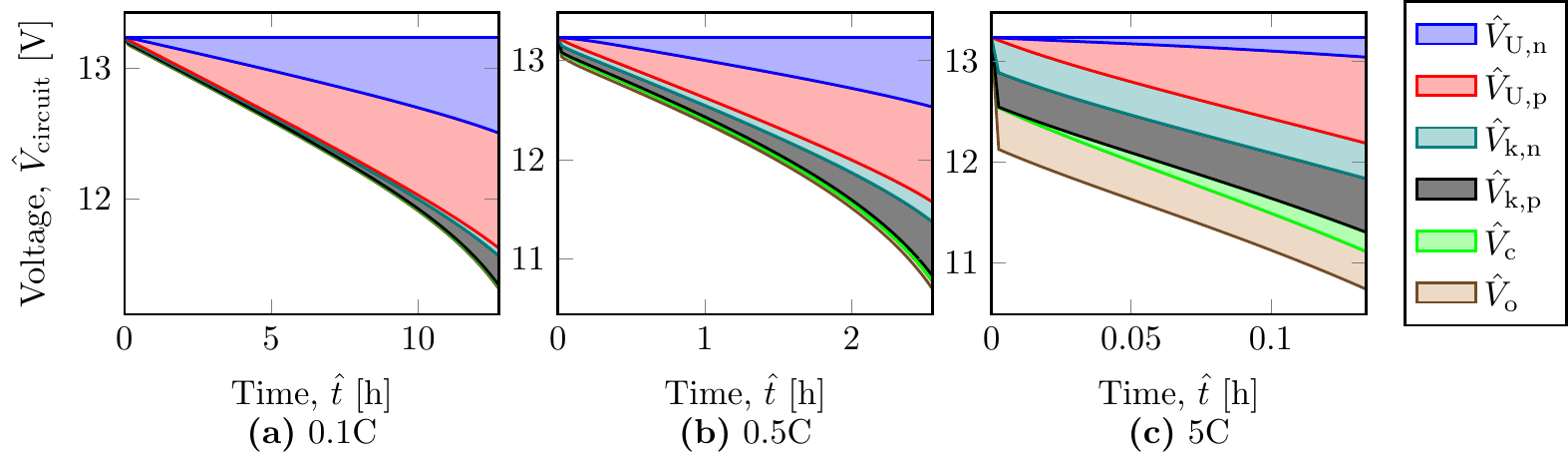}
	\caption{Split of voltage drop into dimensional components, as given by \eqref{eq:voltage_breakdown} and \eqref{eq:voltage_breakdown_details}, for a constant-current discharge using the parameters from literature (Tables~\ref{tab:dimless_params} and \ref{tab:dim_params}).}
	\label{fig:voltage_breakdown}
\end{figure*}
As well as obtaining a faster solution to the model, the composite solution allows us to identify the individual overpotentials that contribute to the total drop in voltage from full charge. We write the total dimensional voltage as the sum of the initial voltage, \begin{equation}
\hat{V}^0 = U\ocpp^\ominus - U\ocpn^\ominus + \frac{RT}{F}\left(U\ocpp(c^0) - U\ocpn(c^0)\right),
\end{equation}
and individual voltage drops~\cite{daigle2013electrochemistry},
\begin{equation}\label{eq:voltage_breakdown}
\hat{V} = \hat{V}^0 + \frac{RT}{F}\left(V_\text{U,n} + V_\text{U,p} + V_\text{k, n} + V_\text{k,p} + V_\text{c} + V_\text{o}\right),
\end{equation}
where $V_{\text{U},i}$, $i=\text{n},\text{p}$ are the open-circuit voltages; $V_{\text{k},i}$, ${i}=\text{n},\text{p}$ are the kinetic overpotentials, accounting for losses due to the reactions at the electrode-electrolyte interfaces; $V_\text{c}$ is the concentration overpotential, accounting for losses due to concentration gradients; and $V_\text{o}$ is the Ohmic overpotential in the electrolyte, accounting for losses due to the electric resistance of the electrolyte. Equation~\ref{eq:voltage_breakdown} would usually include a term to account for Ohmic losses in the solid electrodes, but in our reduced-order models this term is zero since $\iota_\text{s}$ is large (c.f. equation~\ref{eq:Phis_uniform}).

Combining \eqref{eq:V0}, \eqref{eq:Phi1} and \eqref{eq:ODa_algebraic}, we identify
\bse
\begin{align}\label{eq:voltage_breakdown_details}
V_\text{U,n} &= - U\ocpn\left(c\pow{0}\right) - \Cd\,\bar{c}\pow{1}_\text{n}U'\ocpn\left(c\pow{0}\right), \\
V_\text{U,p} &= U\ocpp\left(c\pow{0}\right) + \Cd\,\bar{c}\pow{1}_\text{p} U'\ocpp\left(c\pow{0}\right), \\
V_\text{k,n} &= - \sinh^{-1}\left(\frac{\mathrm{i}_\text{cell}}{2\jecdnz\ell_\text{n}}\right) + \frac{\Cd\,\bar{j}\pow{1}_{0,\text{n}}}{\jecdnz}\tanh\left(\eta_\text{n}\pow{0}\right), \\
V_\text{k,p} &= - \sinh^{-1}\left(\frac{\mathrm{i}_\text{cell}}{2\jecdpz\ell_\text{p}}\right) - \frac{\Cd\,\bar{j}\pow{1}_{0,\text{p}}}{\jecdpz}\tanh\left(\eta_\text{p}\pow{0}\right), \\
V_\text{c} &= \frac{\Cd\,\chi\pow{0}}{c\pow{0}}\left(\bar{c}\pow{1}_\text{p}-\bar{c}\pow{1}_\text{n}\right), \\
V_\text{o} &= -\Cd\,\mathrm{i}_\text{cell}\left(\frac{\ell_\text{n}}{3\keffz_\text{n}}+\frac{\ell_\text{sep}}{\keffz_\text{s}}+\frac{\ell_\text{p}}{3\keffz_\text{p}}\right),
\end{align}
\ese
Together with the quasi-static formulas \eqref{eq:c0} for $c\pow{0}$ and \eqref{eq:c1} for $c\pow{1}$, equations \eqref{eq:voltage_breakdown} and \eqref{eq:voltage_breakdown_details} give an exact formula for the voltage that is valid for most operating C-rates (below 0.5C). For higher C-rates, we must solve \eqref{eq:c_comp} and use \eqref{eq:c1_comp}, instead of \eqref{eq:c1}, to find  $c\pow{1}$.

In Figure~\ref{fig:voltage_breakdown}, we show the relative contribution from each of the terms in \eqref{eq:voltage_breakdown_details}.
At low C-rates (Figure~\ref{fig:voltage_breakdown}a), the drop in voltage is almost entirely due to the change in the OCV of the two electrodes as the concentration changes, $V_\text{U,n}$; 
this is why the LOQS solution is accurate enough in this regime.
As we increase the C-rate (Figure~\ref{fig:compare_voltages}b), the first non-OCV effects to become important are the kinetic overpotentials, $V_\text{k,n}$, 
with the kinetic overpotential in the positive electrode being greater.
Finally, at very high C-rates (Figure~\ref{fig:compare_voltages}c), we see that the sharp initial drop in voltage is due to both kinetic overpotentials and the ohmic overpotential 
in the electrolyte in roughly equal measures.

The breakdown of voltages given by \eqref{eq:voltage_breakdown} could be used to help guide any optimisation of a battery design.
For example, the behaviour observed in Figure~\ref{fig:voltage_breakdown}b suggests that an effective way to improve the power capacity of the battery (by reducing the voltage drop) might be to target reduction of the kinetic overpotential in the positive electrode.
Alternatively, this allows us to explore the trade-off between energy capacity and power capacity of the battery, and hence optimise the behaviour of the battery for specific applications. For  example, increasing the total width, $L$, of an electrode pair would increase the energy capacity of the battery. However, increasing $L$ would also increase the diffusional C-rate (Table~\ref{tab:dimless_params}), and so increase the contributions to the voltage drop in \eqref{eq:voltage_breakdown_details}, decreasing the power capacity of the battery. It becomes evident that we need to consider first-order effects in order to fully understand the trade-off between energy capacity and power capacity; the LOQS model would naively suggest that increasing $L$ is always the optimal strategy.
A third option could be to fix the total width $L$ (\textit{i.e.} fix the energy capacity) and optimise the relative widths of the negative electrode, separator and positive electrode in order to minimise the voltage drop given by \eqref{eq:voltage_breakdown}, and hence maximise the power capacity of the battery.
In all three examples suggested, a quantitative optimisation could be done very efficiently using the formulas given in this paper.

\subsection{Parameter fitting}
\label{sec:fit}

Another potential application for our reduced-order models is in parameter fitting. This is an important challenge when comparing models to experimental data, as many parameters in the model cannot be identified individually \textit{a priori}.
To demonstrate how our model can be useful for parameter fitting, we fit each model to six constant-current discharges of a 17 Ah BBOXX Solar Home battery at intervals of 0.5 A from 3 A to 0.5 A. Each constant-current discharge is followed by a two-hour rest period during which the current is zero.

In order to use as few fitting parameters as possible, we take standard values for all parameters (given in Tables~\ref{tab:dimless_params} and \ref{tab:dim_params}) except for the following:
\begin{enumerate}[label=(\roman*)]
\item the maximum electrode porosities, $\varepsilon_\text{n}^\text{max}$ and $\varepsilon_\text{p}^\text{max}$, which we assume to be equal;
\item the separator porosity, $\varepsilon_\text{sep}^\text{max}$;
\item the exchange current-density for the negative electrode, $j_\text{n}^\text{ref}$ (we assume that ${j_\text{p}^\text{ref} = j_\text{n}^\text{ref}/10}$);
\item a `resistance in the wires', $R_\text{circuit}$, which accounts for ohmic losses outside the battery within the BBOXX Solar Home (we take away $I_\text{circuit}R_\text{circuit}$ from the voltage output by the model before fitting to data).
\item an initial SOC, $q^0$, for five of the curves (we assume that the first discharge, at 3A, starts from full SOC);
\end{enumerate}
This gives a total of nine fitting parameters for six curves.

To fit our model to the data, we minimise the sum-of-squares of the voltage prediction error. We do this in Python with both a derivative-based (\texttt{leastsquares}~\cite{scipy}) or derivative-free (DFO-GN~\cite{cartis2017derivative}) optimisation algorithm.
\begin{figure}
	\centering
	\includegraphics{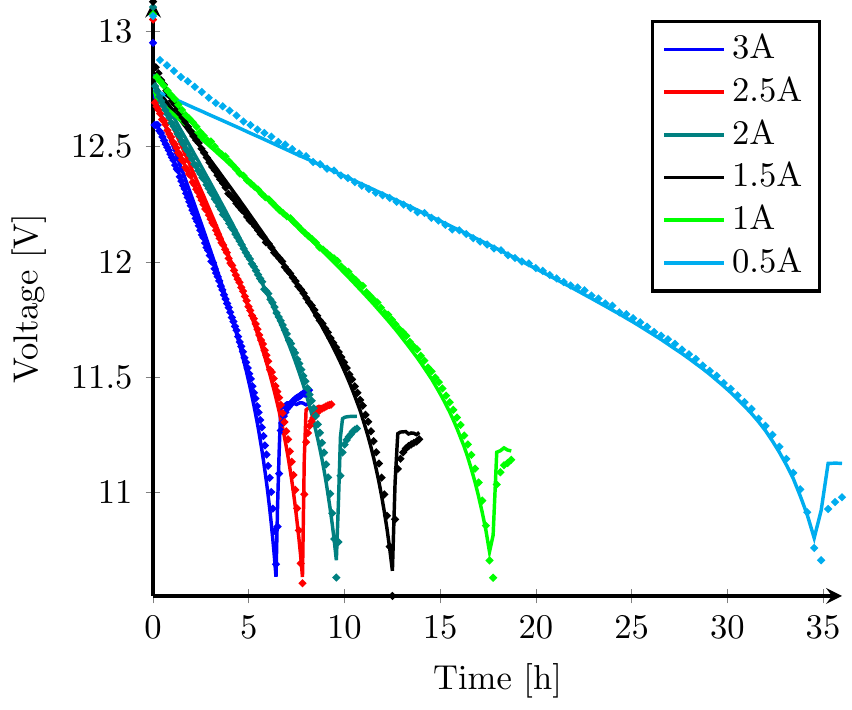}
	\caption{Comparing data (dots) with results from full numerical model (lines) for a range of currents, with parameters fitted using DFO-GN (Tables~\ref{tab:dimless_params} and \ref{tab:dim_params}, except $\varepsilon^\text{max}$, $j^\text{ref}$ and $q^0$ from  Table~\ref{tab:fit_params}).}
	\label{fig:fits}
\end{figure}

Since the highest discharge current is quite low (3A corresponds to a C-rate of 0.18C), both the first-order quasi-static model and composite model agree well with the full numerical model. These three models give a very good fit to the data except near the start of the discharge, and during the short rest period afterwards, as shown in Figure~\ref{fig:fits}. In Table~\ref{tab:fit_performance}, we compare the final sum-of-squares error, as well as the time taken to find the minimum when starting from the parameters from literature of Tables~\ref{tab:dimless_params} and \ref{tab:dim_params}, for each optimisation algorithm/model combination.
As expected, the quasi-static models are faster than the composite model, with the numerical model being by far the slowest. However, this is compounded by the fact that the derivative-based algorithm from SciPy cannot find the best fit for the composite and numerical models, and so we must resort to derivative-free models. Thus optimising the first-order quasi-static model is thirty times faster than optimising the full numerical model.

As shown in Table~\ref{tab:fit_params_selection}, the optimal parameters found by SciPy with the FOQS model are almost the same as those found by DFO-GN with the numerical model, with the exception of the exchange current-density, $j_\text{n}^\text{ref}$. Hence one could first fit all the parameters using SciPy and the FOQS model, and then only need to fit $j_\text{n}^\text{ref}$ with DFO-GN and the numerical model.
\begin{table*}[t]
	\centering
	\begin{tabular}{|c|c|c|c|c|c|c|c|c|}
		\hline
		& \multicolumn{2}{c|}{LOQS} & \multicolumn{2}{c|}{FOQS} & \multicolumn{2}{c|}{Composite} & \multicolumn{2}{c|}{Numerical} \\
		\cline{2-9}
		Algorithm & Error & Time & Error & Time & Error & Time & Error & Time \\
		\hline
		SciPy \texttt{leastsquares} & 14.03 & 36 & 13.43 & 48 & 328.72 & 122 & 351.52 & 1443 \\
		DFO-GN & 14.03 & 74 & 13.43 & 184 & 17.04 & 96 & 9.93 & 1144 \\
		\hline
	\end{tabular}
	\caption{Performance of fitting algorithms for each model. Error is the sum-of-squares of voltage prediction errors, in Volts; time is CPU time in seconds, obtained on an AMD FX(tm)-4350 Quad-Core Processor.}
	\label{tab:fit_performance}
\end{table*}
\begin{table*}[t]
	\centering
	\begin{tabular}{|c|c c|c|c|c c c c c|}
		\hline
		& $\varepsilon_\text{n,p}^\text{max}$ & $\varepsilon_\text{sep}^\text{max}$ & $j_\text{n}^\text{ref}$ & $R_\text{circuit}$ & $q^0_\text{2.5A}$ & $q^0_\text{2A}$ & $q^0_\text{1.5A}$ & $q^0_\text{1A}$ & $q^0_\text{0.5A}$ \\
		\hline
		SciPy + FOQS & 0.55& 0.81& 0.19& 0.08& 1.00& 0.98& 0.95& 0.90& 0.89 \\
		DFO-GN + Numerical & 0.74& 0.55& 0.24& 0.07& 1.00& 0.98& 0.95& 0.90& 0.89 \\
		\hline
	\end{tabular}
	\caption{Parameters obtained from SciPy with the FOQS model, and DFO-GN with the Numerical model. All algorithms given Tables~\ref{tab:dimless_params} and \ref{tab:dim_params} as initial data. A full version of this table is given in \ref{app: fitted}.}
	\label{tab:fit_params_selection}
\end{table*}

The discrepancy at early times may be due to effects from charging: we assume that the discharge starts from equilibrium, but this might not be the case following a charge. The final relaxation timescale is much longer than the two timescales in our model -- the diffusion timescale ($\sim 15$ minutes) and the capacitance timescale ($\sim 5$ seconds) -- which suggests that there is an extra physical effect that we have not considered. In particular, non-uniformities in the $y$- and $z$-dimensions may be important for both early time and relaxation behaviour.

\section{Conclusions}
\label{sec:conc}

The asymptotic methods developed in this paper allow us to simulate a discharge of a lead-acid battery with the low complexity and high speed of equivalent-circuit models, while retaining the accuracy and physical insights of electrochemical models.
Further, these methods give important physical insight into the structure of the problem that is not obvious from numerical solutions of the full model. In particular, we observe that at low C-rates, the voltage drop from open-circuit potential is mainly due to kinetic effects; ohmic and concentration overpotentials are relatively very small. It is also important to note that our models are tunable: for discharges at low C-rate, we should choose the leading-order quasi-static model, while for high C-rates we can use the composite model, which still provides a significant speed-up compared to the full model.

There are many exciting applications for the models resulting from our asymptotic methods. Firstly, these models can be used in Battery Management Systems to replace equivalent-circuit models without introducing the complexity of full electrochemical models. We expect to find significant advantages from doing this, as the parameters will be much more robust to different states of charge than less physically based resistances and capacitances. Secondly, as demonstrated in this paper, the asymptotic models can be used to estimate parameter values, and compare results to experimental data, much more quickly than using the full electrochemical models; we have found that models are often not compared to experimental data due to the prohibitively large cost of the full electrochemical models. Thirdly, since the voltage is given in an explicit form, we can easily perform a parameter sensitivity analysis. Fourthly, by analysing the individual voltage drops from open-circuit potential, we can optimise battery design for specific applications, for example trading off between power and capacity.

An important extension for this work is to apply asymptotic methods to more complex models, for example including side reactions occurring during overcharge, or other long-term degradation mechanisms for lead-acid batteries such as corrosion and irreversible sulfation. We expect the ideas developed here to continue to apply in these more complex settings.

Finally, the ideas explored in this paper can be applied to other battery chemistries. In particular, our leading-order quasi-static model is very similar to the Single-Particle Model for lithium-ion batteries~\cite{dey2014nonlinear, di2010lithium, moura2012adaptive, santhanagopalan2006online, wang2015adaptive}, and our composite model is very similar to the Single-Particle Model with electrolyte~
\cite{han2015simplification, kemper2013extended, moura2017battery, prada2012simplified,
rahimian2013extension, tanim2015temperature}.
However, different chemistries and geometries may lead to different parameter sizes, and hence different distinguished limits.

\section*{Acknowledgements}
\label{sec: ack}

This publication is based on work supported by the EPSRC Centre For Doctoral Training in Industrially Focused Mathematical Modelling (EP/L015803/1) in collaboration with BBOXX. JC, CP, DH and CM acknowledge funding from the Faraday Institution (EP/S003053/1).

\section*{List of symbols}

\noindent\textbf{Variables}
\begin{description}[leftmargin=!, labelwidth=1cm, font=\normalfont]
  \item[$c$] concentration \hfill mol m$^{-3}$
  \item[$\varepsilon$] porosity \hfill -
  \item[$j$] interfacial current density \hfill A m$^{-2}$
  \item[$\bm{i}$] current density (3D) \hfill A m$^{-2}$
  \item[$i$] current density in $x$-direction \hfill A m$^{-2}$
  \item[$\bm{v}$] velocity (3D) \hfill m s$^{-1}$
  \item[$v$] velocity in $x$-direction \hfill m s$^{-1}$
  \item[$\bm{N}$] ion flux (3D) \hfill mol m$^{-2}$ s$^{-1}$
  \item[$p$] pressure \hfill Pa
  \item[$\Phi$] potential \hfill V
\end{description}

\noindent\textbf{Subscripts}
\begin{description}[leftmargin=!, labelwidth=1cm, font=\normalfont]
  \item[n] in negative electrode
  \item[sep] in separator
  \item[p] in positive electrode
  \item[$+$] of cations
  \item[$-$] of anions
  \item[w] of solvent (water)
  \item[e] of electrolyte
  \item[s] of solid (electrodes)
\end{description}

\noindent\textbf{Superscripts}
\begin{description}[leftmargin=!, labelwidth=1cm, font=\normalfont]
  \item[$0$] initial
  \item[max] maximum
  \item[$(0)$] leading-order
  \item[$(1)$] first-order
  \item[eff] effective
  \item[surf] surface
  \item[$\square$] convective
\end{description}

\noindent\textbf{Accents}
\begin{description}[leftmargin=!, labelwidth=1cm, font=\normalfont]
  \item[$\di{}$] dimensional
  \item[$\bar{}$] averaged
\end{description}

\begin{appendix}
\setcounter{table}{0}

\section{Parameters}
\label{sec:params}

The concentration-dependent functions are given in Table~\ref{tab:functions}.

\begin{table*}[t]
\centering
\begin{tabular}{|c|c|c|c|}
\hline
\multicolumn{2}{|c|}{Dimensional} & \multicolumn{2}{c|}{Dimensionless} \\
\hline
$\di{D}\left(\di{c}\right)$
 	& $(1.75+2.6\times10^{-4}\di{c})\times10^{-9}$
	& ${D}(c)$
	& $\di{D}(c^\text{max}c)/\di{D}(c^\text{max})$ \\
$\di{\chi}\left(\di{c}\right)$
 	& $0.49 + 4.1\times10^{-4}\di{c}$
	& ${\chi}(c)$
	& $2(1-t_+^\text{w})\di{\chi}(c^\text{max}c)/(1-\alpha c)$ (\textdagger) \\
$\di{\kappa}\left(\di{c}\right)$
 	& \makecell[l]{$\di{c}\exp\left(6.23 - 1.34\times10^{-4}\di{c}\right.$\\\hspace{2cm}$\left. - 1.61\times10^{-8}\di{c}^2\right)\times10^{-4}$}
	& ${\kappa}(c)$
	& $RT\di{\kappa}(c^\text{max}c)/F^2D^\text{max}c^\text{max}$ \\
$\di{c}_\text{w}\left(\di{c}\right)$
 	& $(1-\di{c}\bar{V}_\text{e})/\bar{V}_\text{w}$
	& ${c}_\text{w}(c)$
	& $\di{c}_\text{w}(c^\text{max}c)/\di{c}_\text{w}(c^\text{max})$\\
$\di{j}_0\left(\di{c}\right)$
 	& $j^{\text{ref}} \left( \frac{\di{c} }{c^{\text{ref}} } \right)^{ \left| \frac{s_+}{n_{\text{e}}} \right| + \left| \frac{s_-}{n_{\text{e}}} \right| } \left( \frac{\di{c}_{\text{w}} }{c_{\text{w}}^{\text{ref}} } \right)^{\left| \frac{s_\text{w}}{n_{\text{e}}} \right| }$
	& ${j}_0(c)$
	& $\mathcal{A}L\di{j}_0(c^\text{max}c)/287\mathcal{C}$ \\
$\di{U}\ocpn\left(\di{c}\right)$
 	& \makecell[l]{$U\ocpn^\ominus - 0.074\log m- 0.030\log^2m$ \\\hspace{.9cm}$- 0.031\log^3m - 0.012\log^4m$ (\textdaggerdbl)}
	& ${U}\ocpn(c)$
	& $\frac{F}{RT}\left(\di{U}\ocpn(c^\text{max}\di{c}) - U\ocpn^\ominus\right)$ \\
$\di{U}\ocpp\left(\di{c}\right)$
 	& \makecell[l]{$U\ocpp^\ominus + 0.074\log m+ 0.033\log^2m$ \\\hspace{.9cm}$+ 0.043\log^3m + 0.022\log^4m$ (\textdaggerdbl)}
	& ${U}\ocpp(c)$
	& $\frac{F}{RT}\left(\di{U}\ocpp(c^\text{max}\di{c}) - U\ocpp^\ominus\right)$ \\
\hline
\end{tabular}
\caption{Functions of concentration, ${c}$. References available in part I, and relevant parameters in Table~\ref{tab:dim_params}.
(\textdagger) ${\alpha = -(2\bar{V}_\text{w} - \bar{V}_\text{e})c^\text{max}}$.
(\textdaggerdbl) ${m(\di{c}) = \di{c}\bar{V}_\text{w}/[(1-\di{c}\bar{V}_\text{e})M_\text{w}]}$.}
\label{tab:functions}
\end{table*}

\renewcommand{\arraystretch}{1}
\begin{table}[t]
\centering
\begin{tabular}{|c|c c c|c|}
\hline
\multirow{2}{*}{Parameter} & \multicolumn{3}{c|}{Value} & \multirow{2}{*}{Units} \\
\cline{2-4}
& n & sep & p & \\
\hline
$L$ & \multicolumn{3}{c|}{$3.65\times 10^{-3}$} & m \\
$A_\text{cs}$ & \multicolumn{3}{c|}{$7.4\times 10^{-3}$} & m$^2$ \\
$s_{+}$ & $-1$ & - & $-3$ & - \\
$s_{-}$ & $1$ & - & $-1$ & - \\
$s_\text{w}$ & $0$ & - & $2$ & - \\
$n_\text{e}$ & $2$ & - & $2$ & - \\
$\bar{V}_\text{w}$ & \multicolumn{3}{c|}{$1.75\times 10^{-5}$} & m$^3$ mol$^{-1}$ \\
$\bar{V}_\text{e}$ & \multicolumn{3}{c|}{$4.50\times 10^{-5}$} & m$^3$ mol$^{-1}$ \\
$M_\text{w}$ & \multicolumn{3}{c|}{$1.8\times 10^{-2}$} & kg mol$^{-1}$ \\
$F$ & \multicolumn{3}{c|}{$96485$} & C mol$^{-1}$ \\
$R$ & \multicolumn{3}{c|}{$8.314$} & J mol$^{-1}$ K$^{-1}$ \\
$T$ & \multicolumn{3}{c|}{$298.15$} & K \\
$t^\text{w}_+$ & \multicolumn{3}{c|}{$0.72$} & - \\
$j^\text{ref}$ & $8\times10^{-2}$ & - & $6\times10^{-3}$ & A m$^{-2}$ \\
$c^\text{max}$ & \multicolumn{3}{c|}{$5.6\times10^{3}$} & mol m$^{-3}$ \\
$\mathcal{A}$ & $2.6\times10^6$ & - & $2.05\times10^7$ & m$^{-1}$ \\
$Q$ & \multicolumn{3}{c|}{$17$} & Ah \\
$U^\ominus$ & -0.295 & - & 1.628 & V \\
\hline
\end{tabular}
\caption{Relevant dimensional parameters from the literature, for Table~\ref{tab:functions}. Parameters with several values indicate different values in negative electrode (n), separator (sep) and positive electrode (p). References are available in part I.}
\label{tab:dim_params}
\end{table}

\section{Concentration in the first-order quasi-static solution}
\label{app:c1}
\setcounter{table}{0}

\bse\label{eq:c1}
\begin{equation}
c\pow{1} = k(t)+\begin{cases}
\tilde{c}\pow{1}\n, &\quad 0<x<\ell_\text{n}, \\
\tilde{c}\pow{1}\s, &\quad \ell_\text{n}<x<1-\ell_\text{p}, \\
\tilde{c}\pow{1}\p, &\quad 1-\ell_\text{p}<x<1.
\end{cases}
\end{equation}
where $k$ is an arbitrary function of $t$ and
\begin{align}
	\tilde{c}\pow{1}\n =& \frac{x^2-\ell_\text{n}^2}{2\Deffz_\text{n}}\left(\od{\left(\varepsilon_\text{n}\pow{0}c\pow{0}\right)}{t} - \frac{s_\text{n}\mathrm{i}_\text{cell}}{\ell_\text{n}}\right), \\
	\tilde{c}\pow{1}\s =& \frac{(x-\ell_\text{n})^2}{2\Deffz_\text{sep}}\varepsilon_\text{sep}^\text{max}\od{c\pow{0}}{t}
	\nonumber \\
	&+ \left(\od{\left(\varepsilon_\text{n}\pow{0}c\pow{0}\right)}{t} - \frac{s_\text{n}\mathrm{i}_\text{cell}}{\ell_\text{n}}\right)\frac{\ell_\text{n}(x-\ell_\text{n})}{\Deffz_\text{n}}
	\\
	\tilde{c}\pow{1}\p =& \frac{(x-1)^2-\ell_\text{p}^2}{2\Deffz_\text{p}}\left(\od{\left(\varepsilon_\text{p}\pow{0}c\pow{0}\right)}{t} + \frac{s_\text{p}\mathrm{i}_\text{cell}}{\ell_\text{p}}\right)
	\nonumber \\
	&+\left(\frac{\varepsilon_\text{sep}^\text{max}\ell_\text{sep}}{2\Deffz_\text{sep}}\od{c\pow{0}}{t}\right.
	\nonumber \\
	&+\left.\frac{\ell_\text{n}}{\Deffz_\text{n}}\od{\left(\varepsilon_\text{n}\pow{0}c\pow{0}\right)}{t} - \frac{s_\text{n}\mathrm{i}_\text{cell}}{\Deffz_\text{n}}\right)\ell_\text{sep}
\end{align}
\ese
We note that the piece-wise quadratic form of \eqref{eq:c1} justifies the assumption of Knauff~\cite{knauff2013kalman}.
To find $k(t)$, we expand \eqref{eq:dcdt} to second-order in $\Cd$:
\begin{multline}\label{eq:dcdt_ODa2}
\pd{}{t}\left(\varepsilon\pow{0}c\pow{1} + \varepsilon\pow{1}c\pow{0}\right) = \pd{}{x}\left( D^\text{eff,(1)}\pd{c\pow{1}}{x}\right)\\
+ \Deffz\pds{c\pow{2}}{x}+ sj\pow{1}.
\end{multline}
Integrating \eqref{eq:dcdt_ODa2} from $x=0$ to $x=1$ and using the fact that ${\partial c\pow{1}/\partial x = \partial c\pow{2}/\partial x = 0}$ at $x=0,1$, together with \eqref{eq:depsdt_ODa} and \eqref{eq:j_BC_ODa}, we find that
\begin{equation}
\pd{}{t}\left(\myint{0}{1}{\varepsilon\pow{0}c\pow{1}}{x}\right) = 0,
\end{equation}
and so, by \eqref{eq:ICs_ODa}, $\myint{0}{1}{\varepsilon\pow{0}c\pow{1}}{x} \equiv 0$. Hence
\begin{equation}\label{eq:k}
k = -\frac{\myint{0}{\elln}{\varepsilon\pow{0}\n\tilde{c}\pow{1}\n}{x}
+ \myint{\elln}{1-\ellp}{\varepsilon\pow{0}\s\tilde{c}\pow{1}\s}{x}
+ \myint{1-\ellp}{1}{\varepsilon\pow{0}\p\tilde{c}\pow{1}\p}{x}}
{\ell_\text{n}\varepsilon_\text{n}\pow{0}
+ \ell_\text{sep}\varepsilon_\text{sep}^\text{max}
+ \ell_\text{p}\varepsilon_\text{p}\pow{0}}
\end{equation}

\section{Transient solution}
\label{app:transient}
\setcounter{table}{0}

Denoting the time of the jump by $t^*$, we rescale time with $\tau = (t-t^*)/\Cd$, and define $C(x, \tau) = c(x, t)$, $E(x, \tau) = \varepsilon(x, \tau)$, $J(x, \tau) = j(x, t)$, $P(x, \tau) = \Phi(x, t)$ and $\mathcal{V}(\tau) = V(t)$. Then \eqref{eq:summary_simplified} becomes
\bse\label{eq:summary_transient}
\begin{align}
\pd{}{\tau}(EC) &= \pd{}{x}\left(\Deff\pd{C}{x}\right) + \Cd\,sJ, \label{eq:dcdt_transient}\\
\pd{E}{\tau} &= -\Cd\,\beta^\text{surf}J, \label{eq:depsdt_transient}\\
\Cd\,{J} &= \pd{}{x}\left[\keff\left(\chi\pd{\ln(C)}{x} - \pd{P}{x}\right)\right], \label{eq:i_transient}\\
J_\text{n} &= 2\Jecdn\sinh\left(-P-U\ocpn(C)\right), \label{eq:jn_transient}\\
J_\text{p} &= 2\Jecdp\sinh\left(\mathcal{V}-P-U\ocpp(C)\right), \label{eq:jp_transient}
\end{align}
\ese
We now expand the variables in powers of $\Cd$, as done in \eqref{eq:Da_expansion}.
Then \eqref{eq:summary_transient} becomes, to leading order in $\Cd$,
\bse\label{eq:transient_O1}
\begin{align}
\pd{}{\tau}&\left(E\pow{0}C\pow{0}\right) = \Deffz\pds{C\pow{0}}{x}, \label{eq:transient_O1_c}\\
\pd{E\pow{0}}{\tau} &= 0, \label{eq:transient_O1_eps}\\
0 &= \pd{}{x}\left[\keffz\left(\chi\pow{0}\pd{\ln\left(c\pow{0}\right)}{x} - \pd{P\pow{0}}{x}\right)\right], \label{eq:transient_O1_Phi}\\
J\pow{0}_\text{n} &= 2\Jecdnz\sinh\left(-P\pow{0}-U\ocpn\left(c\pow{0}\right)\right), \label{eq:transient_O1_jn}
\\
J\pow{0}_\text{p} &= 2\Jecdpz\sinh\left(\mathcal{V}\pow{0}-P\pow{0}-U\ocpp\left(c\pow{0}\right)\right),
\label{eq:transient_O1_jp}
\end{align}
\ese
and to first order in $\Cd$,
\bse\label{eq:transient_ODa}
\begin{align}
&\pd{}{\tau}\left(E\pow{0}C\pow{1}\right) = \Deffz\pds{C\pow{1}}{x} + {sJ\pow{0}} - \pd{}{\tau}\left(C\pow{0}E\pow{1}\right), \label{eq:transient_ODa_c}\\
&\pd{E\pow{1}}{\tau} = -\beta^\text{surf}J\pow{0}, \\
&0 = \pd{}{x}\left[\keffz\left(\frac{\chi\pow{0}}{C\pow{0}}\pd{C\pow{1}}{x} - \pd{P\pow{1}}{x}\right)\right], \\
&J\pow{1}_\text{n} = 2\left(\Jecdno\sinh\left[{H}_\text{n}\pow{0}\right] + \Jecdnz H\pow{1}_\text{n}\cosh\left[H\pow{0}_\text{n}\right]\right) \\
&J\pow{1}_\text{p} = 2\left(\Jecdpo\sinh\left[{H}_\text{p}\pow{0}\right] + \Jecdpz H\pow{1}_\text{p}\cosh\left[H\pow{0}_\text{p}\right]\right)
\end{align}
where
\begin{align}
&H\pow{0}_\text{n} = -P\pow{0}-U\ocpn\left(c\pow{0}\right), \\
&H\pow{0}_\text{p} = \mathcal{V}\pow{0}-P\pow{0}-U\ocpp\left(c\pow{0}\right), \\
&H\pow{1}_\text{n} = -\left({P}\pow{1} + C\pow{1}{U}'\ocpn\left(c\pow{0}\right)\right), \\
&H\pow{1}_\text{p} = {\mathcal{V}\pow{1} - P\pow{1} - C\pow{1}{U}'\ocpp\left(c\pow{0}\right)}.
\end{align}
\ese
The boundary conditions follow from (\ref{eq:summary_O1}f,g) and (\ref{eq:summary_ODa}e,f):
\bse\label{eq:transient_BCs}
\begin{align}
&\pd{C\pow{0}}{x} = \pd{P\pow{0}}{x} = \pd{C\pow{1}}{x} = \pd{P\pow{1}}{x} = 0 \quad \text{ at } x = 0, 1, \label{eq:transient_BCs_flux}\\
&\myint{0}{\ell_\text{n}}{J\pow{0}_\text{n}}{x} = -\myint{1-\ell_\text{p}}{1}{J\pow{0}_\text{p}}{x} = \mathrm{i}_\text{cell}, \label{eq:transient_BCs_jO1}\\
&\myint{0}{\ell_\text{n}}{J\pow{1}_\text{n}}{x} = \myint{1-\ell_\text{p}}{1}{J\pow{1}_\text{p}}{x} = 0,\label{eq:transient_BCs_jODa}
\end{align}
and the initial conditions are given by the states at the jump time $t^*$.
\ese

Equations \eqref{eq:transient_O1} decouple: we first solve \eqref{eq:transient_O1_eps} for $E\pow{0}$, then \eqref{eq:transient_O1_c} with \eqref{eq:transient_BCs_flux} for $C\pow{0}$, then \eqref{eq:transient_O1_Phi} for $P\pow{0}$ up to an additive constant $A$, and finally use \eqref{eq:transient_O1_jn} and \eqref{eq:transient_BCs_jO1} to find $A$ and \eqref{eq:transient_O1_jp} and \eqref{eq:transient_BCs_jODa} to find $\mathcal{V}\pow{0}$.
Similarly, having found the leading-order solution, equations \eqref{eq:transient_ODa} decouple and can be solved in the same order as \eqref{eq:transient_O1}.
Hence we solve \eqref{eq:transient_O1} and \eqref{eq:transient_ODa} more easily than the full system \eqref{eq:summary_transient}.

However, we note that \eqref{eq:transient_O1}, \eqref{eq:transient_ODa} and the boundary conditions \eqref{eq:transient_BCs} imply that $\myint{0}{1}{C\pow{0}}{x}$ is constant, but $\rvert C\pow{1}\rvert$ grows with time. Hence the asymptotic expansion breaks down for long times since, when $\tau \sim \mathcal{O}(1/\Cd)$, $\rvert\text{$\Cd$}\,C\pow{1}\rvert \sim \rvert C\pow{0}\rvert$.

\section{Parameters from fit to data}
\label{app: fitted}
\setcounter{table}{0}

\begin{table*}[ht]
\centering
\begin{tabular}{|c|c c|c|c|c c c c c|}
\hline
& $\varepsilon_{\text{n},\text{p}}^\text{max}$ & $\varepsilon_\text{sep}^\text{max}$ & $j_\text{n}^\text{ref}$ & $R_\text{circuit}$ & $q^0_\text{2.5A}$ & $q^0_\text{2A}$ & $q^0_\text{1.5A}$ & $q^0_\text{1A}$ & $q^0_\text{0.5A}$ \\
\hline
SciPy + LOQS & 0.55& 0.81& 0.19& 0.08& 1.00& 0.98& 0.95& 0.90& 0.89 \\
SciPy + FOQS & 0.55& 0.81& 0.19& 0.08& 1.00& 0.98& 0.95& 0.90& 0.89 \\
SciPy + Composite & 0.60& 0.90& 0.07& 0.13& 1.00& 1.00& 1.00& 0.98& 1.00 \\
SciPy + Numerical & 0.60& 0.90& 0.08& 0.15& 1.00& 1.00& 1.00& 1.00& 1.00 \\
\hline
DFO-GN + LOQS & 0.75& 0.53& 0.19& 0.08& 1.00& 0.98& 0.95& 0.90& 0.89 \\
DFO-GN + FOQS & 0.51& 0.88& 0.32& 0.07& 1.00& 0.98& 0.95& 0.90& 0.89 \\
DFO-GN + Composite & 0.60& 0.78& 0.12& 0.06& 1.00& 0.98& 0.96& 0.90& 0.89 \\
DFO-GN + Numerical & 0.74& 0.55& 0.24& 0.07& 1.00& 0.98& 0.95& 0.90& 0.89 \\
\hline
\end{tabular}
\caption{Parameters obtained from optimisation algorithms/model combinations. All algorithms were given the values in Tables~\ref{tab:dimless_params} and \ref{tab:dim_params} as initial data for the parameters.}
\label{tab:fit_params}
\end{table*}
The full list of parameters obtained from each combination of fitting algorithm (SciPy or DFOGN) and model (LOQS, FOQS, Composite or Numerical) is given in Table~\ref{tab:fit_params}.

\end{appendix}
\bibliographystyle{unsrt}

\end{document}